\documentclass[letterpaper]{JHEP3}
\usepackage{amsmath}
\usepackage{dsfont}
\usepackage{bm}

\newcommand{\eq}{\begin{equation}}
\newcommand{\eeq}{\end{equation}}

\newcommand{\lp}{\left(}
\newcommand{\rp}{\right)}

\newcommand{\al}{\alpha}

\newcommand{\si}{\sigma}

\title{AdS black holes with arbitrary scalar coupling}

\author{Marco M.~Caldarelli\\
Mathematical Sciences and STAG research centre, University of Southampton,\\
Highfield, Southampton SO17 1BJ, United Kingdom\\
E-mail: \email{M.M.Caldarelli-at-soton.ac.uk}}
\author{Christos Charmousis\\
Laboratoire de Physique Th\'eorique (LPT), Univ. Paris-Sud, CNRS UMR 8627,\\
F-91405 Orsay, France\\
Laboratoire de Math\'ematiques et Physique Th\'eorique (LMPT), CNRS UMR 6083,\\
Universit\'e Fran\c{c}ois Rabelais-Tours, France\\
E-mail: \email{christos.charmousis-at-th.u-psud.fr}}
\author{Mokhtar Hassa\"{\i}ne\\
Instituto de Matem\'atica y F\'{\i}sica, Universidad de Talca,\\
Casilla 747, Talca, Chile\\
E-mail: \email{hassaine-at-inst-mat.utalca.cl}}

\abstract{A general class of axionic and electrically charged black holes for  a self-interacting scalar field nonminimally coupled to  Einstein gravity with a
negative cosmological constant is presented. These solutions are the first examples of black holes
with an arbitrary nonminimal coupling $\xi$ in four dimensions. Moreover, due to the presence of two three-forms fields,
the topology of the horizon of these black holes is planar. We discuss some properties of these solutions electing particular values of the nonminimal coupling parameter. A special case arises when $\xi=1/4$, for which the gravitational field is confined in a region close to the event horizon. We also show that these black holes emerge from stealth AdS configurations as the axionic fields are switched on, and that they can be generated through a Kerr-Schild transformation. Finally, in the appendix, we extend these results to arbitrary dimension.}

\begin{document}
\section{Introduction}
One of the main interests of the AdS/CFT correspondence \cite{Maldacena:1997re} lies in the possibility of studying strongly coupled systems by
mapping them into higher-dimensional gravitational models and establishing a dictionary between both theories. Applications range from the study of strongly coupled quark-gluon plasma to condensed matter theory (e.g.~see \cite{Kovchegov:2011ub,Hartnoll:2009sz} for recent reviews), and recently, such ideas have been used in order to gain a better understanding of  superconductors \cite{Horowitz:2010gk,Hartnoll:2008kx,Herzog:2009xv}.
On the gravity side, the study of holographic superconductors involves charged black holes with nontrivial hair and a planar horizon. In order to mimic the nonzero condensate behavior of the superconductor, the black hole is required to develop hair at low temperature that should
disappear at higher temperature via some thermodynamic phase transition. In the simplest case, this will correspond to having a planar and bald AdS-Reissner-Nordstr\"om (AdS RN) black hole at high temperature that spontaneously
generates hair at low temperatures. However, this task is non-trivial and is rendered difficult by various no-hair theorems, see e.g.~\cite{Bekenstein:1998aw}. Nevertheless, scalar fields nonminimally coupled to gravity have proven to be an interesting laboratory in order to avoid such no go theorems. Indeed, as shown independently by Bekenstein \cite{Bekenstein:1974sf} and Bocharova, Bronnikov and Melnikov \cite{Bocharova:1970}, conformal scalar fields nonminimally coupled to Einstein gravity
support geometric black hole configurations. However, the solution suffers from the fact that the scalar field diverges at the horizon \cite{Bocharova:1970,Bekenstein:1975ts} making its physical
interpretation and the problem of its stability a subject of debate \cite{Bronnikov:1978mx,McFadden:2004ni}. A way of circumventing the divergence of the scalar field at the horizon
is to introduce a cosmological constant, whose effect is to precisely push this singularity behind the horizon. In presence of a positive cosmological constant, the topology of the horizon remains unchanged \cite{Martinez:2002ru}, whereas a negative cosmological constant requires the horizon to be hyperbolic \cite{Martinez:2005di}. These solutions have been generalized in \cite{Anabalon:2012tu} by considering a more general potential term. It must be stressed that, in all cases, the nonminimal coupling parameter $\xi$ is always the conformal one in four dimensions, that is $\xi=\frac{1}{6}$, and the horizon topology is either spherical or hyperbolic. However, as shown recently, the black hole potential can be tuned by the addition of $p$-form fields \cite{Bardoux:2012aw}, and, in particular, the event horizon of the  AdS black holes with a conformally coupled scalar can be forced to be planar by introducing a pair of axionic ($3$-form) fields \cite{Bardoux:2012tr}. This partially solves the latter problem although then one can question the meaning of these additional axionic fields persisting even for the bald black holes.

Is the additional conformal symmetry, when $\xi=\frac16$, essential to the construction of these black holes, or does it only make their construction easier? What can one say for nonminimally coupled scalars, with $\xi\neq\frac16$? Up to now their construction has been impossible, at least for spherical and hyperbolic horizon black holes. The purpose of this article is to make progress on this front by presenting planar black hole solutions with an arbitrary value of the nonminimal coupling parameter $\xi$ of the scalar field. To our knowledge\footnote{The only exception being in $2+1$ dimensions, where a BTZ black hole solution with a stealth nonminimally coupled scalar field, with arbitrary coupling $\xi$, has been constructed in \cite{AyonBeato:2004ig}.}, these are the first exact black hole solutions with $\xi\neq\frac16$.

As we shall see, this task may be achieved choosing an appropriate potential for the scalar field, and supplementing the theory with suitable extra fields, following \cite{Bardoux:2012tr}.
Specifically, we consider a self-interacting scalar field nonminimally coupled to Einstein gravity with a negative cosmological constant, together with a standard Maxwell term and two three-form fields nonminimally coupled to the scalar field. The distinctive form of the scalar potential is such that the theory allows for exact AdS solutions with a stealth scalar field for any $\xi$, with the scalar field also being homogeneous in the boundary directions. The metric of such solutions is locally AdS, and a nontrivial scalar field is floating on top of them without backreacting on the geometry. This phenomenon has been known for some time for $2+1$ dimensional AdS gravity \cite{Natsuume:1999cd,AyonBeato:2004ig}, in Minkowski background \cite{AyonBeato:2005tu} and in $D$-dimensional AdS \cite{ABMTZ}. Our starting step is the self-interacting potential presented in \cite{AyonBeato:2005qq,AyonBeato:2006jf} (in the context of pure radiation constraints on AdS wave backgrounds) which allows solutions with Poincar\'e metric and a stealth field \cite{ABMTZ}. Asking in addition that the field is homogeneous in the flat boundary directions relates the coupling constants in the scalar potential\footnote{In the notation of \cite{AyonBeato:2005qq,AyonBeato:2006jf}, this condition reads $\lambda_1=\lambda_2^2$. Here, we define $b=\lambda_2$ and use it as the single coupling constant remaining in $U(\Phi)$ in the rest of this article.}, leaving us with the potential $U(\Phi)$ depending on the couplings $(\xi,b)$\footnote{We are grateful to E.~Ay\'on-Beato, C.~Mart\'\i{}nez, R.~Troncoso and J.~Zanelli for pointing this out and sharing their results \cite{ABMTZ}.}. Finally, we turn on two nonminimally coupled three-form fields in the spirit of \cite{Bardoux:2012aw}, and we find a two-parameter family of exact static solutions, valid for any nonminimal coupling $\xi\neq\frac14$, in which we also allow for a nontrivial Maxwell field.

The general form of these solutions is rather complicated, and a comprehensive analysis is outside the scope of this article and not of actual interest. What we will do instead, is to point out some interesting subfamilies for which the algebraic expression of the metric simplifies, and show that they generically represent well-behaved, locally asymptotically AdS (for $0<\xi<1/4$) spacetimes containing a planar black hole. Not surprisingly, the particular couplings that we will examine actually correspond to conformal couplings in some space-time dimension other than 4. We will furthermore generalize the case $\xi=1/6$ to an additional two parameter family of solutions.

The special value $\xi=\frac14$ of the nonminimal coupling needs a separate treatment, because the self-interacting potential $U(\Phi)$ is singular for that value of the coupling. Only when $b=1$ has it a finite limiting potential, that depends now logarithmically on the scalar field. Again, a couple of $3$-form fields prove providential to construct AdS black holes with a planar horizon. The peculiarity of these solutions is that their gravitational field is localized to a region close to the event horizon, leaving in the asymptotic region only an exponential tail hinting at the presence of the black hole. Interestingly, $\xi=1/4$ is the limiting value of the conformal coupling for a large number of spacetime dimensions $D$, limit for which gravity localizes \cite{Emparan:2013moa}. It would be interesting to investigate this observation further.

The paper is organized as follows. In the next section, we introduce the action of our model, derive the corresponding field equations and present a family of exact solutions. Section 3 is
devoted to the properties of the solutions (black holes nature, event horizon and singularities). We will consider particular nonminimal couplings in order to
discuss the properties of the solutions. In the following section, we will see that the solutions of a scalar field nonminimally coupled with axionic fields can be seen as originated from a stealth configuration given by a scalar field nonminimally coupled to an AdS background. From this perspective, we will see that the axionic fields arise naturally as a good candidate. This will allow us to determine the particular form of the nonminimal coupling of the axionic fields. Finally,
in the last section we address some future extensions of the present work. An appendix is devoted to the extension of these results in arbitrary dimension.

\section{Action, field equations and solutions}
In four dimensions, we consider the following action
\eq
S=\int\! d^4x\sqrt{-g}\left(\frac{R-2\Lambda}{16\pi G}-\frac{\epsilon(\Phi)}{12}
\sum_{i=1}^{2}H_{\mu\nu\rho}^{(i)}H^{(i)\mu\nu\rho}-\frac{1}{2}\partial_{\mu}\Phi\partial^{\mu}\Phi-\frac{\xi}{2}R\Phi^2-U(\Phi)-
\frac{F_{\mu\nu}F^{\mu\nu}}{16\pi}\right),
\label{actioncharged}\eeq
where the constant $\Lambda=-3/\ell^2$ represents the negative cosmological constant, $F_{\mu\nu}$ stands for the Maxwell strength, $\Phi$ is a scalar field
nonminimally coupled with gravity through the nonminimal coupling parameter $\xi$, and $U(\Phi)$ is a potential that depends on the scalar field, whose expression reads
\begin{eqnarray}
U(\Phi)=\frac{\Phi^2}{8\pi G l^2}\Big[3\xi (1-6\xi)+\frac{2\xi^2}{(1-4\xi)^2} \Big(2(1-6\xi)+b \Phi^\frac{1-4\xi}{2\xi}\Big)^2\Big],
\label{potxi}
\end{eqnarray}
with $b$ a coupling constant. Note that when $b=0$, we have simply a $\Phi^2$ mass potential, with the notable exception of conformal coupling occurring at $\xi=1/6$. In addition, we have two exact three-forms
\eq
{\cal H}^{(i)}=\frac{1}{3!}H^{(i)}_{\mu\nu\rho}dx^\mu\wedge dx^\nu\wedge dx^\rho,\qquad i=1,2
\eeq
originating from two
Kalb-Ramond potentials. These forms are nonminimally coupled to the scalar field through the coupling function $\epsilon$ given by
\begin{eqnarray}
\epsilon(\Phi)=\frac{(1-8\pi G\xi \Phi^2)^2}{f(\sqrt{8\pi G}\Phi)},
\label{epsilonphi}
\end{eqnarray}
where we have defined the function $f(x)$ depending on two coupling constants $\sigma_1$ and $\sigma_2$ by
\begin{eqnarray}
f(x)=&& (1-6\xi)(1-4\xi)(1-\xi x^2)^2+2\xi \left(1-4\xi-\xi x^2\right)
\nonumber\\
&&
-\,\sigma_1\xi^2\left(\frac{1}{1-4\xi}-\xi x^2\right)x^{\frac{1}{\xi}-2}
+16\sigma_2\xi^3x^\frac{1}{2\xi}.
\label{epsilon}\end{eqnarray}
We will see in a later section how the choice of this potential comes about.

The field equations obtained by varying the action with respect to the different dynamical fields yield
\begin{alignat}{2}
&G_{\alpha\beta}-\frac{3}{\ell^2}g_{\alpha\beta}=8\pi G\,T_{\alpha\beta},&\qquad
&\nabla_\mu\left(\epsilon(\Phi) H^{(i)\mu\alpha\beta}\right)=0,\\
&\Box\Phi=\xi R\,\Phi+\frac{\mathrm{d}U(\Phi)}{\mathrm{d}\Phi}+
\frac{1}{12}\frac{d\epsilon(\Phi)}{d\Phi}\sum_{i=1}^{2}H_{\alpha\beta\gamma}^{(i)}H^{(i)\alpha\beta\gamma},&\qquad
&\nabla_{\mu}F^{\mu\nu}=0,
\end{alignat}
where the corresponding energy-momentum tensor is defined by
\begin{eqnarray}
T_{\alpha\beta}&=&\nabla_{\alpha}\Phi\nabla_{\beta}\Phi
-g_{\alpha\beta}\left(\frac{1}{2}g^{\mu\nu}\nabla_{\mu}\Phi
\nabla_{\nu}\Phi+U(\Phi)\right)
+\xi\left(g_{\alpha\beta}\Box-\nabla_{\alpha}\nabla_{\beta}
+G_{\alpha\beta}\right)\Phi^2\\
&&+\,\epsilon(\Phi)\sum_{i=1}^{2}\left(\frac{1}{2}H_{\alpha\mu\nu}^{(i)}H_{\beta}^{(i)\mu\nu}-\frac{1}{12}g_{\alpha\beta}H_{\mu\nu\rho}^{(i)}H^{(i)\mu\nu\rho}\right)
+\frac{1}{4\pi}\left(F_{\alpha\gamma}F_{\beta}^{\,\,\,\gamma}-\frac{1}{4}g_{\alpha\beta}F_{\mu\nu}F^{\mu\nu}\right).\nonumber \label{eq:emtcharged}
\end{eqnarray}
It is clear from these different expressions that the special case $\xi=\frac{1}{4}$ must be treated separately; this will be done below. Hence, for a nonminimal coupling parameter $\xi\not=\frac{1}{4}$, the theory is completely determined by the four coupling constants $(\xi,b,\si_1,\si_2)$, and an exact solution is given by
\begin{alignat}{2}
&ds^2=-F(r)dt^2+\frac{dr^2}{F(r)}+\frac{r^2}{\ell^2}(d x_1^2+dx_2^2),&\qquad
&\Phi(r)=\frac1{\sqrt{8\pi G}}(ar+b)^{\frac{-2\xi}{1-4\xi}}\nonumber\\
&{\cal H}^{(i)}=\frac{ p}{\sqrt{8\pi G}\,(1-4\xi)^2\ell^2 \epsilon(\Phi)}\, dt \wedge dr \wedge d x^i,&\qquad
 &{\cal F}=-\frac{ q}{r^2}dt\wedge dr.
\label{solution}\end{alignat}
The lapse function $F(r)$  and the constant $ a$ are given by
\eq
F(r)=\frac{r^2}{\ell^2}-\frac{ p^2(ar+b)^{\frac{4\xi}{1-4\xi}}}{(ar+b)^{\frac{4\xi}{1-4\xi}}-\xi}\left(1+\frac{2G\mu}{r}-\frac{G q^2}{ p^2r^2}\right),\quad
a=\frac1{G\mu}\lp b-\sigma_2\rp.
\label{sol4}
\eeq
This family of solutions is completely determined by the three integration constants $(\mu, p, q)$ related to mass, electric and axionic charges respectively, and subject to the constraint
\eq
q^2=\frac{ p^2\mu^2G}{(b-\sigma_2)^2}\lp
2b\sigma_2-b^2-\frac{\sigma_1}{2(1-4\xi)}\rp,
\label{q^2}
\eeq
hence effectively yielding a two-parameter family of solutions. The constant $(r,t)$ sections are flat.

Finally, we note that the $\mu\rightarrow0$ limit of this solution is finite if $0<\xi<1/4$, and gives a black hole solution with vanishing scalar and Maxwell fields, and lapse function $F(r)=r^2/\ell^2-p^2$, first obtained in \cite{Bardoux:2012aw}.
Also, choosing
\eq
\xi=\frac16,\qquad
b = \sqrt{\frac{2\al\ell^2}{8\pi G}},\qquad
\si_1=-\frac1{9},\qquad
\si_2=0,
\label{confcase}\eeq
we recover precisely the conformally coupled solution of \cite{Bardoux:2012tr}, with exactly the same notation.

\paragraph{Special value $\xi=\frac{1}{4}$:}
The previous expressions clearly make no sense when $\xi=\frac{1}{4}$. Indeed, to find solutions for this particular value of the coupling, the potential $U(\Phi)$ as well as the $\epsilon(\Phi)$ function have to be modified to acquire a logarithmic dependence on the scalar field. They are given by
\begin{eqnarray}
U(\Phi)=\frac{1}{32\pi G\ell^2}\Phi^2\left(3+2\ln\left(\frac{\Phi}{b}\right)^2+6\ln\left(\frac{\Phi}{b}\right)\right),
\label{pot14}
\end{eqnarray}
where $b$ is a constant, and
\begin{eqnarray}
\epsilon(\Phi)=\frac{\left(8 \pi G \Phi^{2}-4\right)^{2}}{f(\sqrt{8 \pi G}\Phi)},
\end{eqnarray}
with
\eq
f(\Phi)=2\Phi^{2}\ln\left(\frac{\Phi}{b}\right)\left[4\ln\left(\frac{\Phi}{b}\right)-\Phi^{2}\right]
        +4\left(\Phi^2-4\right)+\sigma_{1}\Phi^{2}\left[{8}\ln\left(\frac{\Phi}{b}\right)-\left(\Phi^{2}-4\right)\right].
\eeq
For this theory, a solution is given by
\begin{alignat}{2}
& ds^2=-F(r)dt^2+\frac{dr^2}{F(r)}+\frac{r^2}{\ell^2}(dx^2_1+dx^2_2),&\qquad
&\Phi(r)=\frac{b\,e^{ar}}{\sqrt{8\pi G}},\nonumber\\
&{\cal H}^{(i)}=\frac{p}{2\sqrt{8\pi G}\,\ell^2\epsilon(\Phi)}\, dt \wedge dr \wedge dx^i,&\qquad
&{\cal F}=0,
\label{solution14}\end{alignat}
where the lapse function and the constant $a$ take the form
\eq
F(r)=\frac{r^{2}}{\ell^{2}}-\frac{p^{2}}{b^{2} e^{2 a r}-4}
        \left(1+\frac{\mu}{r}\right),\qquad a=\frac{\sigma_{1}-1}{\mu}.
\label{lapse14}\eeq
Note that, in contrast with the other values of $\xi$, this solution cannot accommodate an electrically charged Maxwell field, and we always have ${\cal F}=0$. Hence the theory is fully determined by two couplings $(b,\sigma_1)$, and the solutions are determined by the two integration constants $(\mu,p)$.

\paragraph{Double scaling limits:}
It is worth mentioning that simpler but different families of solutions can be obtained as double scaling limits of the previous configurations. When $\xi\neq1/4$, we can take the limit $\sigma_2\rightarrow b$, $\mu\rightarrow0$, keeping the ratio $a=(b-\si_2)/G\mu$ fixed. The resulting configuration is finite, solves the equations of motion with coupling $\si_2=b$, and consists in the fields \eqref{solution} with lapse function and constraint given by
\eq
F(r)=\frac{r^2}{\ell^2}-\frac{ p^2(ar+b)^{\frac{4\xi}{1-4\xi}}}{(ar+b)^{\frac{4\xi}{1-4\xi}}-\xi}\left(1-\frac{G q^2}{ p^2r^2}\right),\qquad
q^2=\frac{ p^2}{Ga^2}\lp b^2-\frac{\sigma_1}{2(1-4\xi)}\rp.
\label{dsl}\eeq
This family of solutions is completely determined by the two integration constants $(a, p)$.

A similar double scaling limit can be taken when $\xi=1/4$. It is obtained from the solution \eqref{solution14}-\eqref{lapse14} taking the limit $\sigma_1\rightarrow0$, $\mu\rightarrow0$ while keeping the ratio $a$ fixed. The resulting family of solutions is again given by \eqref{solution14}, together with
\eq
F(r)=\frac{r^{2}}{\ell^{2}}-\frac{p^{2}}{b^{2} e^{2 a r}-4},
\label{dsl14}\eeq
and the parameter $a$, in addition to $p$, being now an arbitrary integration constant.

\section{Analysis of the solutions}

In this section, we will provide an analysis of the solutions obtained in the previous section. Our
study is not exhaustive because of the complexity of the different expressions involved to describe the solutions. We will start by giving  some general comments concerning the black hole nature of the solutions and the location of their event horizon. We will also discuss, in more detail, special values of nonminimal coupling parameter, which would correspond to the conformal coupling in $D=3$, $4$, $5$ and $D=6$ dimensions (i.e.~$\xi=\frac{1}{8}$, $\frac{1}{6}$,
$\frac{3}{16}$ and $\xi=\frac{1}{5}$), as well as the minimal case $\xi=0$ and the
 case $\xi=\frac{1}{2}$, which corresponds to a non decaying linear scalar field. The last subsection is devoted to the particular case $\xi=\frac{1}{4}$, for which the potentials in the theory become logarithmic and where only uncharged electrical solutions will be presented. We will assume $\xi\in\left[0,\frac14\right]$ throughout the section, with the only exception of the $\xi=\frac12$ case.

\subsection{General comments}
First of all, turning off the axionic fields by setting $p=0$, the solutions reduce to AdS stealth configurations: indeed the lapse functions \eqref{sol4} and \eqref{lapse14} become $F(r)=r^2/\ell^2$, and the geometry is that of AdS, in Poincar\'e coordinates\footnote{This generalizes the observation made in \cite{Bardoux:2012tr} for the special case \eqref{confcase} to arbitrary values of the couplings.}. The scalar field $\Phi(r)$ keeps a nontrivial radial profile, but nevertheless does not backreact on the metric. Also, the constraint \eqref{q^2} requires the Maxwell field to vanish on these stealth configurations. Therefore, we can understand the solutions \eqref{solution} and \eqref{solution14} as the result of turning on the two $3$-form fields in a gravitational stealth background. In many cases, as we will presently show, these $3$-forms will also induce a planar event horizon, leading to regular, locally AdS black hole geometries.

Condition \eqref{q^2} obviously constrains the parameters in our action. The integration constant  $q$ is real if and only if
\eq
|b-\sigma_2|\leq \sqrt{\sigma_2^2-\frac{\sigma_1}{2(1-4\xi)}}.
\eeq
In particular, $q=0$ whenever the above bound is saturated namely at
\eq
b= \sigma_2 \pm \sqrt{\sigma_2^2-\frac{\sigma_1}{2(1-4\xi)}}.
\label{boundb}\eeq
In order for the scalar field to decay at infinity we have that $0<\xi<1/4$. We will restrict ourselves to these values of $\xi$ and will only examine the extra case $\xi=1/2$ which gives us a linear scalar field diverging at infinity in section~\ref{sec:linear}. This latter case will be treated as a paradigm of a non-decaying scalar. Suppose then that the scalar does decay and is therefore a decreasing function for large enough $r$. We start by looking at the asymptotes of the lapse function $F(r)$, which can be conveniently written as
\eq
F(r)=\frac{r^2}{\ell^2}-\frac{ p^2 \Phi^{-2}}{\Phi^{-2}-8\pi G\xi}\left(1+\frac{2G\mu}{r}-\frac{G q^2}{ p^2r^2}\right).
\eeq
Clearly, as $r\rightarrow \infty$, $F(r)$ diverges quadratically, and the spacetime is asymptotically locally AdS. What are the possible singularities for finite $r$? To begin with, the scalar field explodes at $r_\Phi=- \frac{G \mu b}{b-\sigma_2}$. This is the type of singularity encountered in the BBMB solution \cite{Bekenstein:1974sf,Bocharova:1970}. The metric is not singular there but the scalar field is. Furthermore, the effective Newton constant\footnote{Defined by $G_{\sf eff}=G/(1-8\pi G\xi\Phi^2)$.} diverges for $r=r_N$ defined by $\Phi^{-2}(r_N)=8\pi G \xi$. This is the denominator in $F(r)$ and will yield a genuine spacetime singularity. A notable exception happens for the conformal case $\xi=1/6$ with $\sigma_1=-1/9$ and $\sigma_2=0$, see \eqref{confcase}. In this case, the singularity in $r=r_N$ due to the denominator vanishing cancels out with the last term in the parenthesis of $F(r)$, thus providing a regular solution at this point for the conformal frame \cite{Bardoux:2012tr}, rendering it special and better behaved in general.
The singularity, absent in the metric in the conformal frame, however, reappears in the minimal frame as the metric conformal factor is zero there. Therefore the apparent cancelation is in fact a red herring and one must be careful in order to interpret this solution as a genuine black hole. Only when this curvature singularity is hidden by the event horizon\footnote{We call $r_+$ the largest root of $F(r)$.} $r_N<r_+$, we have a legitimate black hole. In the conformal frame this is reflected in the black hole entropy naively becoming negative when $r_+<r_N$, despite the spacetime being regular.
Either way, in the generic case, the singularity in $r=r_N$ is present in the general expression for $F(r)$. Last but not least there is the usual black hole singularity at $r=0$. Now we determine which of these singularities we hit first. This will be the endpoint of spacetime for each case. First of all, given the bounds on $\xi$, we have that $r_\Phi<r_N$. Hence suppose first that $r_\Phi>0$, i.e.~that the scalar explodes before hitting $r=0$. Then $r=r_N$ is the spacetime singularity, and the endpoint of spacetime. Given the asymptotic behavior at infinity,  if $F(r)\rightarrow -\infty$ at $r=r_N$, then we have at least one zero of $F(r)$ and hence an event horizon cloaking the singularity. This is true if in a neighborhood of $r=r_N$ we have that
\eq
1+\frac{2G\mu}{r}-\frac{G q^2}{ p^2r^2}> 0,
\eeq
which translates into the condition
\eq
\label{cond1}
\sigma_1> 2(1-4\xi)\lp2\xi^{\frac{1-4\xi}{4\xi}}\sigma_2-\xi^{\frac{1-4\xi}{2\xi}}\rp.
\eeq
In the limiting case when we have the equality in the previous relation, a cancelation occurs in $F(r)$, similar to the one we observed for $\xi=1/6$, $\sigma_1=-1/9$ and $\sigma_2=0$. When this relation is satisfied, the metric is regular in $r=r_N$.
For the conformal $\xi=1/6$ coupling, this happens when $\sigma_1=-\frac19+\frac{2\sqrt6}9\sigma_2$. Hence, whenever $r_N\geq 0$ and the above condition \eqref{cond1} is fulfilled, we have a black hole solution with at least one event horizon. The situation is more delicate when $r_N<0$. Then for $ q\neq 0$ the lapse function $F(r)\rightarrow +\infty$ much like the RN solution and only for big enough mass parameter we will get a black hole with this time at least two horizons. Obtaining the condition for this to happen is straightforward but algebraically tedious due also to the additional condition \eqref{q^2}, and we will not examine it here.

In the subsections that follow we will in particular examine two cases: the case $b=0$ and secondly the case in which the ``common root condition'' is fulfilled. The latter case will be possible when we have integral powers appearing in \eqref{sol4}. Then we will ask that the numerator and denominator in $F$ within the last parenthesis \eqref{sol4} have a common root.
For the former case we can already get a useful overall picture of the solutions setting $b=0$. Several interesting things happen then. For a start, the scalar field explodes at the geometric singularity $r=r_\Phi=0$. Also, the scalar potential $U(\Phi)$ is a $\Phi^2$ mass term, except when $\xi=1/6$ for which case we have no potential at all. Finally, $q$ real imposes that $\sigma_1<0$, whereas $\sigma_2 \neq 0$.
Lastly, we note that the lapse function becomes
\eq
F(r)=\frac{1}{1-8\pi G\xi\Phi^{2}}\left[\frac{r^2}{\ell^2}-p^2 \left(1+\frac{2G\mu}{r}-\frac{G q^2}{ p^2r^2}\right)-\frac{\xi}{\ell^2}a^\frac{-4\xi}{1-4\xi}r^{2-\frac{4\xi}{1-4\xi}}\right],
\label{b=0}
\eeq
and that it behaves in a similar way to the one of the hyperbolic AdS RN black hole\footnote{Note that the metric in the minimal frame is given by $\tilde g_{\mu\nu}=\lp1-8\pi G\xi\Phi^2\rp g_{\mu\nu}$, and the overall, monotonic conformal factor is absent from the lapse function \eqref{b=0} in that frame. Thus, the horizon structure in the minimal frame is set by the term in square brackets of \eqref{b=0}.}, with the addition of the last term due to the scalar field secondary hair. For $\xi=1/2$ this term dominates at large $r$ over the cosmological constant term. For $\xi=1/8$ we get a linear term in $r$. Then, for $\xi=1/6$ we get a constant term affecting the horizon curvature just like the axions.  Last but not least, for $\xi=3/16$ and $\xi=1/5$ we get an effective mass and charge term respectively. In fact, note that setting $\xi=\xi_D$ with
\eq
\xi_D=\frac{D-2}{4(D-1)},
\label{xi}\eeq
we pick up the cases of conformal coupling in $D$ spacetime dimensions. Naturally, since our solution is four-dimensional, our scalar field $\Phi$ is conformally coupled only for $\xi=1/6$, i.e.~$\xi=\xi_4$. Note, however, that for these values of $\xi$ the exponents appearing in $F(r)$  \eqref{sol4} are precisely $\frac{4\xi}{1-4\xi}=D-2$. These integer powers single out the $\xi=\xi_D$ cases as special and make the analysis of the solution far easier. As the value of $\xi$ increases, the scalar field decays more slowly at infinity. It is interesting to note that the values $\xi=1/8$, $1/6$, $3/{16}$, and $\xi=1/5$ correspond to the conformal coupling in 3, 4, 5 and 6 spacetime dimensions respectively. Also, $\xi=1/4$ is the value of the conformal coupling \eqref{xi} in the limit of infinitely many spacetime dimensions. We can now attack these special cases in turn.

\subsection{The conformal coupling case}
The scalar field is conformally coupled for the value of $\xi=1/6$ in four dimensions. In order for $q$ to be real we have
\eq
|b-\sigma_2|\leq \sqrt{\sigma_2^2-\frac{3\sigma_1}{2}},
\label{confb}\eeq
and the lapse function takes the form
\eq
F(r)=\frac{r^2}{\ell^2}-p^2\lp 1+\frac{G\mu b}{r(b-\sigma_2)}\rp^2\lp 1-\frac{\frac{2G\mu \sigma_2}{b-\sigma_2}\Big[r+\frac{G\mu}{b-\sigma_2}(b-\frac{1}{12 \sigma_2}-\frac{3\sigma_1}{4\sigma_2})\Big]}{\lp r+\frac{G\mu b}{b-\sigma_2} \rp^2-\frac{G^2\mu^2}{6(b-\sigma_2)^2}} \rp,
\eeq
for generic $\sigma_{1,2}$ and $b$. In the previous section, we saw how the solution behaves for $b=0$. The only case with a known black hole solution corresponds to putting $\sigma_1=-1/9$ and $\sigma_2=0$ but $b\neq 0$ \cite{Bardoux:2012tr}. This is when a simplifying cancelation occurs in $F(r)$. In order to see this and to generalize this solution we ask that the numerator and denominator in $F$ within the last parenthesis have a common root. As we saw earlier, this is achieved for
\eq
\sigma_1=-\frac{1}{9}\pm \frac{2\sqrt6}{9} \sigma_2,
\eeq
and we then get,
\eq
F(r)=\frac{r^2}{\ell^2}-p^2\lp 1+\frac{G\mu b}{r(b-\sigma_2)}\rp^2 \lp 1-\frac{\frac{2G\mu \sigma_2}{b-\sigma_2}}{r+\frac{G\mu}{b-\sigma_2}(b\pm\frac{1}{\sqrt{6}})} \rp.
\eeq
This is the common root condition and we will apply it extensively for each particular coupling{\footnote{A similar simplification occurs for the coupling potential $\epsilon(\Phi)$.}}. Note now that indeed if we take additionally $\sigma_2=0$ we get the previously found black hole solution that extends the MTZ black hole to the case of a planar horizon \cite{Bardoux:2012tr}. From here stem two particular cases which both have $q=0$ given by \eqref{boundb}, namely $b=\mp \frac{1}{\sqrt{6}}$ and $b=\mp \frac{1}{\sqrt{6}}+2\sigma_2$. In fact note that the permitted range of $b$, defined by \eqref{confb}, is in-between these values. The former value gives the lapse function,
\eq
F(r)=\frac{r^2}{\ell^2}-p^2 \lp 1+\frac{2G\mu}{r} + \frac{G^2 \mu^2 b}{(b-\sigma_2)r^2} -\frac{2 G^3 \mu^3 b^2 \sigma_2}{(b-\sigma_2)^3r^3}\rp,
\eeq
and the solution is locally asymptotically AdS. Note that although $q=0$ the metric still behaves as if it were electrically charged. Additionally however, we pick up a term in $1/r^3$. Taking $\sigma_2=\mp\frac{1}{2\sqrt{6}}$ gives us a perfect cube. The other value yields a more complicated lapse function, but still describes a black hole.

Finally, let us suppose now that there is no potential for the scalar field, i.e.~we set $b=0$ and hence $\sigma_1\leq 0$ in order for $q^2\geq 0$. The solution takes the form
\eq
F(r)=\lp1-\frac{G^2 \mu^2}{6\sigma_2^2r^2}\rp^{-1}\left[ \frac{r^2}{\ell^2}-\left(\frac{G^2 \mu^2}{6\ell^2 \sigma_2^2} +p^2\right) -\frac{2p^2G \mu}{r}-\frac{3G^2\sigma_1 p^2\mu^2}{2\sigma_2^2 r^2}\right].
\eeq
It is easy to see that $F(r)$ is that of AdS RN with a constant term\footnote{Also dubbed {\em curvature} term, since it corresponds to the curvature of the transverse space \cite{Vanzo:1997gw}.} that depends on $\mu$, $p$ and $\sigma_2$. This term, in absence of matter fields, must vanish for a planar black hole \cite{Lemos:1994xp,Huang:1995zb}, but allowing for the three-form fields it can be made negative \cite{Bardoux:2012aw}. Interestingly, tuning the coupling $\si_2$ of the three-form, we can make it positive for these new black holes, effectively producing a planar AdS black hole with a lapse function mimicking the one of a spherical black hole.

\subsection{Case of $\xi=1/8$}
In this case the value of $\xi$ corresponds to the conformal coupling in 3 dimensions. Here, although the coupling is not conformal in four dimensional spacetime, the lapse function $F(r)$ simplifies considerably,
\eq
F(r)=\frac{r^2}{\ell^2}-\frac{ p^2\lp1+\frac{G\mu b}{r(b-\sigma_2)}\rp}{r\lp r+\frac{G\mu }{b-\sigma_2}(b-\frac{1}{8})\rp}\left(r^2+2G\mu r-\frac{G q^2}{ p^2}\right),
\eeq
where $q$ is well defined if and only if
\eq
|b-\sigma_2|\leq \sqrt{\sigma_2^2-\sigma_1}.
\eeq
When the inequality is saturated we have $q=0$. The common root condition is obtained for $4\sigma_1=\sigma_2-\frac{1}{16}$. The solution then simplifies to,
\eq
F(r)=\frac{r^2}{\ell^2}-p^2\lp 1+\frac{2G \mu (b-\sigma_2+\frac{1}{16})}{r(b-\sigma_2)} +\frac{G^2 \mu^2 b(b-2\sigma_2+\frac{1}{8})}{r^2 (b-\sigma_2)^2}\rp.
\eeq
The lapse function is that of a hyperbolic RN AdS black hole for the relevant mass and charge which depend on the theory parameters. Note in particular that setting $q=0$ does not mean the absence of a charge-like term for the black hole.
 Here also note that the coupling function $\epsilon(\Phi)$, given in \eqref{epsilon}, simplifies to
\eq
\epsilon(\Phi)=\frac{4}{\frac{1}{8}\lp\sigma_2-\frac{1}{16}\rp\Phi^2+1}.
\eeq
The coupling $\epsilon(\Phi)$ is constant for $\sigma_2=1/16$, which gives us,
\eq
F(r)=\frac{r^2}{\ell^2}-p^2\lp 1+\frac{G \mu b}{r(b-\frac{1}{16})} \rp^2.
\eeq
This is the  $\xi=1/8$ version of the conformal case studied in \cite{Bardoux:2012tr}.

The other interesting case is to take $b=0$ upon which we have that $\sigma_1\leq 0$ and that our potential $U(\Phi)\sim \Phi^2$ is a mass term. The lapse function can be written with ease from \eqref{b=0} and corresponds to a quartic in $r$. Imposing simply $4\sigma_1=\sigma_2-\frac{1}{16}$, we get
\eq
F(r)=\frac{r^2}{\ell^2}-p^2-\frac{2 G \mu }{r}\lp1-\frac{1}{16\sigma_2}\rp,
\eeq
the lapse function of the hyperbolic AdS black hole \cite{Mann:1996gj,Vanzo:1997gw}.

\subsection{Cases $\xi=3/16$ and $\xi=1/5$}
When $\xi$ takes the above values we have the conformal coupling for five- and six-dimensional spacetime respectively. The general expression is complicated and we will just consider a $U(\Phi)\sim \Phi^2$ potential i.e.~we will restrict ourselves to $b=0$.
When $\xi=3/16$ the lapse function takes the form,
\eq
F(r)=\frac{1}{1+\frac{3}{16} \lp\frac{G \mu}{\sigma_2 r}\rp^3}\lp \frac{r^2}{\ell^2}-p^2-\frac{G \mu}{r}\lp2 p^2-\frac{3 G^2 \mu^2}{16 \sigma_2^3}\rp -\frac{2p^2G^2 \mu^2 \sigma_1}{\sigma_2 r^2}\rp.
\eeq
For $\frac{G \mu}{\sigma_2}\geq 0$ this solution has the same horizon structure as the hyperbolic version of AdS RN with the relevant mass and charge which get shifted around according to our theory. When $\xi=1/5$ the solution takes a similar form,
\eq
F(r)=\frac{1}{1-\frac{1}{5} \lp\frac{G \mu}{\sigma_2 r}\rp^4}\lp \frac{r^2}{\ell^2}-p^2-\frac{2G\mu p^2}{r} -\lp \frac{G\mu}{\sigma_2 r}\rp ^2\lp \frac{G^2\mu^2}{5\sigma_2^2 l^2}+\frac{5 p^2 \sigma_1}{2} \rp\rp.
\eeq
Again, this looks very similar to the hyperbolic AdS RN but here we have to be slightly careful since the singularity occurs at finite $r$. In fact we have to make sure that the numerator remains negative as we approach this singularity. Another difference here is that it is the charge term which is shifted by the couplings rather than the mass term in the former case.

\subsection{Minimal coupling to gravity $\xi=0$}
When gravity is minimally coupled we notice that the three-forms are also minimally coupled and the potential $U(\Phi)$ is trivial. The scalar field is a constant that can be set to zero without loss of generality (since it is massless). The solution we then obtain is that of the planar axionic black hole found in \cite{Bardoux:2012aw}.

\subsection{Case of a linear scalar field, $\xi=\frac12$}\label{sec:linear}
Up to now we discussed only cases where the scalar field decays at infinity. We will consider now the case where the scalar field is linear. Since our theory in nonminimally coupled this case may still present some interest. Anyway, it is a typical prototype of a diverging scalar solution at infinity and we wish to investigate its characteristics in order to have a full picture of the solutions for arbitrary $\xi$. We have that $\xi=1/2$. The charge $q$ is well defined if and only if, $|b-\sigma_2|\leq \sqrt{\sigma_2^2+\frac{\sigma_1}{2}}$.
Proceeding as before we ask for a common root condition and obtain $\sigma_1=4(1\pm \sqrt{2}\sigma_2)$.
The lapse function reads,
\eq
F(r)=\frac{r^2}{\ell^2}+ \frac{2}{r^2}\lp\frac{pG\mu}{b-\sigma_2}\rp^2 \frac{r+\frac{\mu G}{b-\sigma_2}\lp b\mp\sqrt{2}-2\sigma_2\rp}{r+\frac{\mu G}{b-\sigma_2}\lp b\mp\sqrt{2}\rp}.
\eeq
Note then when $\sigma_2=0$ there is a naked singularity since $F$ is strictly positive. To remedy this we keep $\sigma_2\neq 0$. Let s take here for simplicity the case where the above inequality is saturated, $b=\pm\sqrt{2}$. We then have $q=0$, and the metric lapse function is that of an asymptotically locally AdS  black hole,
\eq
F(r)=\frac{r^2}{\ell^2}+\frac{2p^2}{r^2}\lp\frac{G\mu}{b-\sigma_2} \rp^2- \frac{4\sigma_2p^2}{r^3} \lp \frac{G\mu}{b-\sigma_2}\rp^3.
\eeq

\subsection{Special value $\xi=\frac{1}{4}$}
This case can be easily studied exhaustively, and presents a new exotic behavior.
First, if $\sigma_1=1$ the scalar field is constant and the metric has lapse function
\eq
F(r)=\frac{r^2}{\ell^2}-\frac{p^2}{b^2-4}-\frac{p^2}{b^2-4}\frac\mu r,
\eeq
and is singular for $|b|<2$, but when $|b|>2$ it represents a planar AdS black hole, with the lapse of an hyperbolic black hole. In this respect it behaves similarly to the planar black holes with $3$-form fields presented in \cite{Bardoux:2012aw}.

The situation is more intriguing when $a>0$, that is for $\sigma_1>1$ and $\mu>0$\footnote{An alternative possibility for $a>0$ is to have $\sigma_1<1$ and $\mu<0$. This case is slightly more complex, and does not add much to the discussion, so we will ignore it in this article.}:
in this case, the metric in the asymptotic region differs from the Poincar\'e metric by exponentially small corrections only.
Also, the function $F(r)$ has a positive root when $|b|>2$, and the spacetime therefore contains a black hole.
Hence, the solution is exponentially close to the AdS ground state in the asymptotic region of the black hole, with the gravitational field confined in a region of size set by $\mu/(\sigma_1-1)$ around the event horizon. In particular, the ADM mass of this black hole vanishes.
Finally, note that the value of the scalar field diverges exponentially in the asymptotic region;  this should however not be a cause of concern, since it behaves as a stealth field for the AdS metric, and  does not backreact on the metric. Somewhat more surprisingly, the scalar field is also such that it shields the $3$-form field from the metric, by cancelling, up to exponentially small terms, all the stress tensor components of the $3$-form fields stress tensor, leaving the asymptotic Poincar\'e region unperturbed.

\section{Derivation of the action and the solution}\label{sec:derivation}
The scalar potential $U(\Phi)$, as well as of the function $\epsilon(\Phi)$ dictating the coupling of the axions to the scalar $\Phi$, have rather complicated expressions, and thus seem unnatural. To understand how they where chosen, and how they lead to the general solutions \eqref{solution} and \eqref{solution14}, recall that we observed in the previous section that these solutions are axionic excitations on top of an AdS stealth background. As explained in the introduction, the most general scalar potential allowing for such configurations is the one presented in \cite{AyonBeato:2005qq}. Moreover, since we want to construct planar black holes, we need to start from a background configuration sharing the same symmetries: the stealth scalar field $\Phi$ can thus depend on the radial Poincar\'e coordinate $r$ only. This further constrains the potential $U(\Phi)$  to assume the form \eqref{potxi} \cite{ABMTZ}.

Next, we need to add the two $3$-form fields to the action, engineering the coupling $\epsilon(\Phi)$ to the scalar field so that the integrability properties are maintained.
For simplicity, we will only consider the electrically neutral case; the addition of the Maxwell term to the action being trivial. Starting from a stealth configuration on the AdS background given by a scalar field nonminimally coupled with a self-interaction potential precisely given by (\ref{potxi}).  By definition, the AdS stealth configuration has a nontrivial scalar field whose energy-momentum tensor evaluated on an AdS background identically vanishes,
\begin{eqnarray}
\label{stealtheqs}
&&G_{\alpha\beta}+\Lambda g_{\alpha\beta}=T^{\sf stealth}_{\alpha\beta}=0,\qquad
\Box\Phi=\xi R\,\Phi+\frac{\mathrm{d}U(\Phi)}{\mathrm{d}\Phi},\\
&&T^{\sf stealth}_{\alpha\beta}=\nabla_{\alpha}\Phi\nabla_{\beta}\Phi
-g_{\alpha\beta}\left(\frac{1}{2}\nabla_{\gamma}\Phi
\nabla^{\gamma}\Phi+U(\Phi)\right)
+\xi\left(g_{\alpha\beta}\Box-\nabla_{\alpha}\nabla_{\beta}
+G_{\alpha\beta}\right)\Phi^2,\nonumber
\end{eqnarray}
where the potential $U(\Phi)$ is given by (\ref{potxi}). The static stealth configuration in four dimensions with planar base manifold reads \cite{ABMTZ}
\begin{eqnarray}
ds^2_0=-r^2dt^2+\frac{dr^2}{r^2}+r^2(dx_1^2+dx_2^2),\qquad \Phi(r)=\left(ar+b\right)^{\frac{-2\xi}{1-4\xi}}.
\label{stealthads}
\end{eqnarray}
Note that the stealth equations (\ref{stealtheqs}) can be viewed as a particular solution of the equations associated to the variation of the following action
\begin{eqnarray}
S^{\sf stealth}=\int d^4x\sqrt{-g}\Big[\frac{R-2\Lambda}{16\pi G}-\frac{1}{2}\partial_{\mu}\Phi\partial^{\mu}\Phi-\frac{\xi}{2}R\Phi^2-U(\Phi)\Big].
\label{stealthaction}
\end{eqnarray}

Let us now operate a Kerr-Schild transformation on the Poincar\'e metric (\ref{stealthads}), with a null and geodesic vector given by $l=dt-\frac{dr}{r^{2}}$.
The transformed metric $\bar g_{\mu\nu}$ is
\eq
d\bar{s}^2=ds^2_0+Mr^{2}F_1(r)\,l\otimes l.
\eeq
Redefining the time coordinate according to
\eq
dt\to dt-\frac{Mr^{2}F_1(r)}{r^{4}(1-MF_1(r))}dr,
\eeq
the metric becomes
\eq
d\bar{s}^2=-r^{2}\Big(1-MF_1(r)\Big)dt^2+\frac{dr^2}{r^2\Big(1-MF_1(r)\Big)}+r^2\left( dx_1^2+dx_2^2\right).
\label{metricBH}
\eeq
In these different expressions, $M$ stands for a constant and $F_1(r)$ is a metric function to be determined.

The idea is now to see what
kind of extra matter can act as a source of the metric background (\ref{metricBH}) in order to solve  Einstein equations\footnote{ The bar notation is used to stress that the quantities are computed with respect to the transformed metric (\ref{metricBH}). }
\begin{eqnarray}
\bar{G}_{\alpha\beta}+\Lambda \bar{g}_{\alpha\beta}=\bar{T}^{\sf stealth}_{\alpha\beta}+\bar{T}^{\sf extra}_{\alpha\beta},
\label{eqsextra}
\end{eqnarray}
where $\bar T^{\sf extra}_{\alpha\beta}$ will correspond to the stress tensor associated to the extra matter source. In order to answer this question, it is interesting to appreciate the effects of the Kerr-Schild transformation on the gravitational equations. Defining
\eq
E_{\alpha\beta}=G_{\alpha\beta}+\Lambda g_{\alpha\beta}-T^{\sf stealth}_{\alpha\beta},
\eeq
we have that $E_{\alpha\beta}$ vanishes for a stealth solution, but when evaluated on the Kerr-Schild metric \eqref{metricBH}, it is easy to prove that
\eq
\bar{E}_{tt}=\Sigma \,\bar{g}_{tt},\qquad\qquad \bar{E}_{rr}=\Sigma\, \bar{g}_{rr},\qquad\qquad \bar{E}_{ii}=\sigma \,\bar{g}_{ii},
\label{rel1}\eeq
where $\Sigma$ and $\sigma$ are functions of the radial coordinate $r$, $F_1$, $F_1^{\prime}$ and $F_1^{\prime\prime}$. All we need to do, to solve the gravitational field equations \eqref{eqsextra}, is to add extra fields to the action, such that on-shell their combined stress tensor $\bar T_{\alpha\beta}^{\sf extra}$ evaluates to $\bar E_{\alpha\beta}$ on the Kerr-Schild metric \eqref{metricBH}, compensating thus precisely these extra terms. This problem has been thoroughly analyzed in \cite{Bardoux:2012aw} with free fields, where it was shown that the problem can be solved introducing one extra three-form field strength for each spacelike direction on the horizon. Turning on only the electric component with one single leg along horizon directions, and distributing these legs isotropically on the horizon, one obtains an aggregated stress tensor with the same algebraic structure as $\bar E_{\alpha\beta}$. Here we see that, despite the presence of interaction, we can apply the same trick and produce the desired stress tensor.

This works as follows. We add to the starting action (\ref{stealthaction}) an extra kinetic term for two axionic fields $H^{(1)}$ and $H^{(2)}$, one for each independent direction on a four-dimensional planar horizon,
\eq
S^{\sf extra}=-\int\! d^4x\sqrt{-\bar{g}}\left(\frac{\epsilon(\Phi)}{12}
\sum_{i=1}^{2}H_{abc}^{(i)}H^{(i)abc}\right),
\eeq
with $\epsilon(\Phi)$ defining their coupling to the scalar. The corresponding stress tensor is
\eq
\bar{T}^{\sf extra}_{\alpha\beta}=\epsilon(\Phi)\sum_{i=1}^{2}\left(\frac{1}{2}H_{\alpha bc}^{(i)}H_{\beta}^{(i)bc}-\frac{1}{12}\bar{g}_{\alpha\beta}H_{abc}^{(i)}H^{(i)abc}\right),
\eeq
while the field equation that follow for the $3$-forms are given by
\eq
\bar{\nabla}_{\alpha}\left(\epsilon(\Phi) {\cal H}^{(i)\alpha\beta\gamma}\right)=0.
\label{pformseq}\eeq
Next, we need to switch on the $3$-forms without breaking the Poincar\'e symmetry of the AdS boundary. This is achieved by allowing them to generate a purely electric field, with single non vanishing components
${\cal H}^{(1)}_{trx_1}$ and ${\cal H}^{(2)}_{trx_2}$. Imposing the equation of motion (\ref{pformseq}), and matching the integration constants to ensure the isotropy of the resulting stress tensor, we obtain
\eq
{\cal H}^{(i)}=\frac{p}{\epsilon}dt\wedge dr\wedge dx^{i},\qquad i=1,2,
\eeq
where $p$ is an integration constant. These fields finally produce a total stress tensor of the form required to solve Einstein's equations,
\eq
\bar{T}_{tt}^{\sf extra}=-\frac{p^2}{\epsilon r^2}\bar{g}_{tt},\qquad
\bar{T}_{rr}^{\sf extra}=-\frac{p^2}{\epsilon r^2}\bar{g}_{rr},\qquad
\bar{T}_{ii}^{\sf extra}=0.
\eeq
Combining these relations with those of (\ref{rel1}), we see that in order to obtain a solution
of the Einstein equations (\ref{eqsextra}) in the background (\ref{metricBH}), the function $\sigma$ must vanish while $\epsilon=-\frac{p^2}{\Sigma r^2}$. The first condition
$\sigma=0$ yields a second-order differential equation for the metric function $F_1$ whose integration allows to determine $\Sigma$. Finally, expressing the radial coordinate
in term of the scalar field (\ref{stealthads}) as
$$
r=\frac{\Phi^{\frac{4\xi-1}{2\xi}}-b}{a}
$$
permits to express the function $\epsilon$ as a function of the scalar field $\Phi$, giving equation (\ref{epsilonphi}).

This explains how the special $3$-form coupling to the scalar was imposed by the requirement of having planar black holes. The latter can be seen to emerge from stealth solutions when the axionic fields are switched on, and can be generated through a Kerr-Schild transformation.

\section{Conclusions}
In this paper, we have considered a four-dimensional action for a scalar field nonminimally coupled to gravity and axionic fields.
We have exhibited, for the first time, a general class of field configurations solving the equations of motion for arbitrary  values of the nonminimal coupling, which we noted $\xi$.
It consists in a two-parameter family of solutions including mass, electric and axionic charge, and a relation in between them. For $0\leq\xi\leq\frac14$, these solutions are asymptotically locally AdS, and contain a planar black hole for appropriate ranges of the parameters.
Our theory has, apart from $\xi$, three more couplings parameterizing the relevant scalar potential and axionic coupling in the action.

We have shown that these solutions can be viewed as the response of an AdS stealth background configuration to the inclusion of axionic charges, and they can be generated through a Kerr-Schild transformation. The stealth configurations develop a planar horizon when the $3$-form fields are switched on, yielding regular black hole geometries. It would be interesting to understand the role of these axionic fields better, and their relation to gravitational stealth configurations. The thermodynamic proprieties of the above family should also be investigated  in order to better understand their relevant phases and consider potential holographic applications. In order to go further in this direction, one needs however to understand first the role played by the axions in the holographic picture, for they are essential in order to get planar horizons.

We should note that one can write the solutions in the minimal frame with ease since the minimally coupled metric is simply given by $\tilde g_{\mu\nu}=\lp1-8\pi G\xi\Phi^2\rp g_{\mu\nu}$. The action itself however, and in particular the conformally transformed scalar field, are in general not given analytically. We found that higher or lower dimensional conformal couplings simplify the relevant solutions. Apart from the $4$-dimensional conformal coupling, we studied  the relevant $\xi$ couplings associated to $3$, $5$, and $6$ dimensions. It may be that the solutions presented here can then be uplifted to higher dimensions. Moreover, the peculiar forms of the scalar potential $U(\Phi)$ and the scalar-axion coupling $\epsilon(\Phi)$ are very special in that they allow exact black hole solutions. This integrability property, in light of the simplifications arising for the above values of $\xi$, hints to a higher-dimensional origin of these potentials, possibly through some generalized dimensional reduction \cite{Kanitscheider:2009as,Gouteraux:2011qh}. Likewise, it is enthralling to entertain the idea that the $\xi=\frac14=\xi_{\infty}$ case, for which the gravitational field is confined to a region close to the horizon, might be linked to the large $D$ limit of some gravitational theory \cite{Emparan:2013moa}. We believe that it should be possible to obtain a deeper understanding of the structure of the action \eqref{actioncharged}, and that it would be fruitful to pursue the research in this direction.

The solutions we have found have a rich horizon structure with Cauchy and event horizons. Generically they present a horizon structure similar to that of a charged AdS hyperbolic black hole. This results from the axionic fields since their charge generically gives a negative curvature term in the lapse function. In the case where the potential is absent and $\xi=1/6$, or $U\sim \Phi^2$ for any other $\xi$, there is an additional term appearing in the lapse function with a variable role, of mass, charge, curvature, depending on the non minimal coupling $\xi$ \eqref{b=0}. In particular for the case of conformal coupling $\xi=1/6$, the extra term plays the role of horizon curvature in the lapse function, allowing even for a positive rather than a negative curvature term. One then may question the existence of a de~Sitter rather than an anti-de~Sitter black hole with a planar topology\footnote{All the solutions presented here are valid for a positive cosmological constant too performing the change $\ell\rightarrow i\ell$.}. However, closer investigation shows that, although we can find a regular static region delimited by two horizons, the solution is never asymptoting de Sitter space.

\begin{acknowledgments}
We thank Eloy Ay\'on-Beato, Yannis Bardoux, Moises Bravo, \'Oscar Dias, Roberto Emparan and Kostas Skenderis for useful discussions.
MC acknowledges support from a grant of the John Templeton Foundation. The opinions expressed in this publication are those of the authors and do not necessarily reflect the views of the John Templeton Foundation.
CC was partially supported by the ANR grant STR-COSMO, ANR-09-BLAN-0157.
MH was partially supported by grant 1130423 from FONDECYT, by grant ACT 56 from CONICYT. CC and MH acknowledge support from CONICYT, Departamento de Relaciones Internacionales ``Programa Regional MATHAMSUD 13 MATH-05''. Finally, MH thanks the LPT, Univ. Paris-Sud in Orsay (Paris)  where part of this project was initiated for their kind hospitality.

\end{acknowledgments}

\appendix
\section{Extension in arbitrary dimension}
In this appendix, we generalize the solutions above to arbitrary dimension, omitting the Maxwell term for simplicity. Also, we set the constants $8\pi G$ and $\ell$ to unity.  Following the same procedure as in four dimensions, we start from stealth AdS solutions, with self-interacting scalar potential $U(\Phi)$ presented in \cite{AyonBeato:2006jf}, and choosing the scalar-axion coupling $\epsilon(\Phi)$ in such a way that static, exact solutions with Poincar\'e symmetry in the boundary directions exist \cite{ABMTZ}. The resulting action is
\eq
\begin{split}
S=\int\! d^Dx\,\sqrt{-g}\left(\frac{R-2\Lambda}{2}\right.&-\frac{1}{2}\partial_{\mu}\Phi\partial^{\mu}\Phi
-\frac{\xi}{2}R\Phi^2-U(\Phi)\\
&\left.-\frac{\epsilon(\Phi)}{2(D-1)!}\sum_{i=1}^{D-2}
{\cal H}^{(i)}_{\alpha_1\cdots \alpha_{D-1}}{\cal H}^{(i)\alpha_1\cdots \alpha_{D-1}}\right),
\end{split}
\eeq
where $\Lambda=-(D-1)(D-2)/2$ is the negative cosmological constant, and where we have introduced $(D-2)$ fields which are exact $(D-1)$-forms ${\cal H}^{(i)}$. The function $\epsilon(\Phi)$ depends on the scalar field $\Phi$ as
\begin{equation}
        \epsilon(\Phi)=\frac{\left(D-2\right)^{2}\left(1-4\,\xi\right)^{2}\left(1-\xi \Phi^{2}\right)^{\frac{D}{D-2}}}{G(\Phi)},
\label{epsilonD}\end{equation}
with
\eq
\begin{split}
G(\Phi)=& {4b\,\xi^{2}\,\Phi^{\frac{1-2\xi}{\xi}}}
\left[8\left(D-2\right)\left(D-3\right)\left(\xi-\xi_{D-2}\right)\Phi^{\frac{4\xi-1}{2\xi}}
        -8\,\xi\left(D-1\right)\left(D-4\right)
        \left(\xi-\xi_{D}\right)\Phi^{\frac{8\xi-1}{2\xi}}
        \right.\\
&\qquad\qquad\qquad\left.\vphantom{\Phi^{\frac{8\xi-1}{2\xi}}}
        +b\,\xi\,(D-2-8\xi)\Phi^{2}-b\left(D-2\right)\right]
        +\left(D-3\right)\left(D-2\right)^{2} \left(1-4\xi\right)^{2}\\
&\quad+4\xi\left[4\left(D-1\right) \xi \left(\xi-\xi_{D}\right)\left(\left(D-2\right)^{2}
        \left(\xi-\xi_{D-1}\right)-2\xi\right)\Phi^{4}
        \right.\\
&\qquad\qquad\left.\vphantom{\left(\xi-\xi_{D}\right)\left(\left(D-2\right)^{2}\right)}
        -\left(D-2\right)\left(2\left(D-1\right)\left(D-3\right)\left(4\xi^{2}+\xi_{D}\right)-\xi\left(4\,D^{2}-18\,D+17\right)\right)\Phi^{2}\right],
\end{split}\nonumber\eeq
and the potential term is again the potential associated to the stealth configuration on the AdS background, and is given by \cite{ABMTZ}
\begin{eqnarray}
U(\Phi)=\frac{\xi}{{(1-4\xi)^2}}\,\left[2\,\xi\,b^{2}\Phi^{\frac{1-2\xi}{\xi}}-8\,(D-1)\left(\xi-\xi_{D}\right)
        \left(2\,\xi\,b\,\Phi^{\frac{1}{2\xi}}-D \left(\xi-\xi_{D+1}\right)\,\Phi^{2} \right)\right],
\end{eqnarray}
where $\xi_D$, given in equation \eqref{xi}, denotes the conformal coupling in $D$ dimensions.
The field equations read
\begin{subequations}
\begin{eqnarray}
&&G_{\mu\nu}-\frac{(D-1)(D-2)}{2}g_{\mu\nu}=\partial_{\mu}\Phi\,\partial_{\nu}\Phi
-g_{\mu\nu}(\frac{1}{2}\partial_{\sigma}\Phi\,\partial^{\sigma}\Phi+U)+\xi\left(g_{\mu\nu}\Box-\nabla_{\mu}\nabla_{\nu}
+G_{\mu\nu}\right)\Phi^2\nonumber\\
&&+\epsilon\sum_{i=1}^{D-2}\Big[\frac{1}{(D-2)!}
{\cal H}^{(i)}_{\mu\alpha_1\cdots \alpha_{D-2}}
{\cal H}_{\nu}^{(i)\alpha_1\cdots \alpha_{D-2}}-\frac{g_{\mu\nu}}{2(D-1)!}{\cal H}^{(i)}_{\alpha_1\cdots \alpha_{D-1}}
{\cal H}^{(i)\alpha_1\cdots \alpha_{D-1}}\Big],\\
\nonumber\\
&&\Box\Phi=\xi R\Phi+\frac{dU}{d\Phi}+\frac{1}{2(D-1)!}\frac{d\epsilon}{d\Phi}\sum_{i=1}^{D-2}{\cal H}^{(i)}_{\alpha_1\cdots \alpha_{D-1}}
{\cal H}^{(i)\alpha_1\cdots \alpha_{D-1}},\\
\nonumber\\
&&\nabla_{\mu}\left(\epsilon{\cal H}^{(i)\mu\alpha_1\cdots\alpha_{D-2}}\right)=0,
\end{eqnarray}
\label{eqs}
\end{subequations}
and a solution is given by
\begin{eqnarray} \label{solaxfield}
       ds^{2}&=&-F(r)\,dt^{2}
        +\frac{1}{F(r)}\,{dr^{2}}+{r^{2}}
        \sum_{i=1}^{D-2} dx_{i}^{2},\qquad \Phi(r)=\left(ar+b\right)^{\frac{2\xi}{4\xi-1}},
        \nonumber\\
        \mathcal{H}^{(i)}&=&\frac{p}{\epsilon(\Phi)} r^{D-4} dt\wedge
        dr \wedge \ldots \wedge dx^{i-1} \wedge dx^{i+1} \wedge
        \ldots \wedge dx^{D-2}
\end{eqnarray}
where the metric function is
\begin{equation}
F(r)=r^{2}- \frac{p^{2}}{\left(1-\xi \left(ar+b\right)^{\frac{4\xi}{4\xi-1}}\right)^{\frac{2}{D-2}}},
\label{solMoises}\end{equation}
with $p$ and $a$ two arbitrary integration constants.

These solutions are the analogue of the double scaling limit solutions \eqref{dsl}, and indeed they reduce to those solutions (with $q=0$) when $D=4$. The reason why we obtain this limiting solution only, effectively with $\mu=0$, and we do not obtain the more general solution corresponding to the family \eqref{solution}-\eqref{sol4}, comes from the fact that the coupling $\epsilon(\Phi)$ between the scalar and $p$-form fields, given by \eqref{epsilonD}, lacks the coupling constants $\si_1$ and $\si_2$ present in \eqref{epsilonphi}. More precisely, when $D=4$ it reduces to the four-dimensional function \eqref{epsilonphi} with the special values $\sigma_1=2(1-4\xi)b^2$ and $\sigma_2=b$ of the couplings.

It is possible to reintroduce these two coupling constants in the $D$-dimensional function $\epsilon(\Phi)$, at the cost of increasing further the complexity of the expressions in this appendix. Reintroducing $\si_1$ would allow to obtain solutions electrically charged under the Maxwell field, while an arbitrary coupling $\si_2$ would push the mass parameter $\mu$ away from the zero value of the double scaling limit. On the whole, one would obtain the generalization to arbitrary dimension of the solution given in \eqref{solution}-\eqref{sol4}.

The reason why this cannot be achieved in closed form can be explained as follows. Starting from the stealth configuration in arbitrary dimension, and operating a Kerr-Schild transformation as explained in Section~\ref{sec:derivation}, the extra components of the energy-momentum tensor associated to the axionic fields read on-shell
\eq
{T}_{tt}^{\sf extra}=
        -\frac{p^{2}\,(D-2)}{2\,\epsilon(\Phi)
        r^2}\,{g}_{tt}, \quad
        {T}_{rr}^{\sf extra} =
        -\frac{p^{2}\,(D-2)}{2\,\epsilon(\Phi)
        r^2}\,{g}_{rr},\quad
        {T}_{ii}^{\sf extra} =
        -\frac{p^{2}\,(D-4)}{2\,\epsilon(\Phi)
        r^2}\,{g}_{ii}.
\eeq
Now in dimension $D\not=4$, the components along the planar direction do not vanish, ${T}_{ii}^{extra} \not=0$. This complicates further the differential equation for the lapse function, leading to an exact solution in closed form only when we take $\mu=0$. 

In more details, defining
\eq
F(r)=r^2\left(1-p^2f(r)\right),\qquad\mbox{with}\qquad  f(r)=\frac{1+h(r)}{r^{2}\left(1-\xi
\left(ar + b\right)^{\frac{4\xi}{4\xi-1}}\right)^{\frac{2}{D-2}}},
\eeq
and considering the combination $(D-4) E_{t}^{t}-(D-2)E_{x_{i}}^{x_{i}}=0$ of Einstein's equations (here we define $E_{\alpha\beta}$ to be the components of Einstein equations, as we did in section~\ref{sec:derivation}) one yields to the following differential equation for the unknown metric function $h(r)$,
\eq
\left( 4\,\xi-1 \right)  \left( D-2 \right)  \left( \xi\, \left( ar+b
\right) ^{{\frac {4\xi}{4\,\xi-1}}}-1 \right)  \left(  h'' r+ h' \left( D-2 \right)  \right)
+4\,a  {\xi}^{2} r \left(D-4 \right)  \left( ar+b \right) ^{ \frac{1}{ 4\,\xi-1
}}h'=0.\\
\label{maseq}
\eeq
In four dimensions, this equation reduces to $rh^{\prime\prime}+2h^{\prime}=0$, and its integration gives precisely the neutral version of the four-dimensional solution (\ref{sol4}).
On the other hand, for dimensions $D\not=4$, this differential equation cannot be integrated in full generality. However, it is clear that equation (\ref{maseq}) admits the trivial solution $h(r)=\text{const}$. This constant can be absorbed without loss of generality into $p^2$. The resulting expression corresponds  to the solution given by (\ref{solMoises}) and leads to the associated function $\epsilon(\Phi)$ given in \eqref{epsilonD}.

More generally, for $D\not=4$, the formal solution of the equation (\ref{maseq}) is given by
\eq
h(r)=C_{1} \int r^{-D+2} \left(\xi(ar+b)^{\frac{4\xi}{4\xi-1}}-1\right)^{-\frac{D-4}{D-2}}dr,
\eeq
and operating the change of variable $x=\xi \left(ar+b\right)^{\frac{4\xi}{4\xi-1}}$ as done in \cite{AyonBeato:2005qq}, the metric function $h(x)$ satisfies the following equation,
\begin{align}
        &(D-2)\xi x (x-1)  \left( -x \left( {
        \frac {x}{\xi}} \right) ^{-\frac{1}{4{\xi}}}+b\xi \right)
        h_{,xx}\nonumber\\
        &\quad
        -\left\{\frac{1}{4}\left[(D-3) \left( D(4\xi-1) +2 \right) x^{2}
        - (D-2)  \left(D(4\xi-1)-8\xi+3\right)x\right]  \left(\frac {x}{\xi}\right)^{-\frac{1}{4\xi}}
        \right.
        \nonumber\\
        &\qquad\qquad\qquad\qquad\qquad\qquad\qquad\left.
        -b \left[
        \left(4(D-4)\xi+D-2\right)x
        -(D-2)\right] \xi
        \vphantom{\left(\frac {x}{\xi}\right)^{-\frac{1}{4\xi}}}\right\}h_{,x}=0.
\end{align}
Nevertheless, there is another particular case for which an exact solution in closed form is possible. If the constant $b$ is set to zero, the last equation reduces to an hypergeometric one,
\eq
 x(x-1) h_{,xx}+{\frac { (D-3)\left(
        D \left( 4\,\xi-1 \right) +2 \right)x- \left( D-2
        \right)  \left( D \left( 4\,\xi-1 \right) -8\,\xi+3 \right)
        }{ 4 \left( D-2 \right) \xi}} h_{,x}=0,
\eeq
whose solution reads
\eq
h(x)=C_1+C_2x^{\frac{(D-3)(1-4\xi)}{4\xi}}{}_2F_1\left(\frac{D-4}{D-2},\frac{(D-3)(1-4\xi)}{4\xi},\frac{D(1-4\xi)+16\xi-3}{4\xi};x\right),
\eeq
with $C_1$ and $C_2$ two integration constants. In this case, we recover a two-parameter family of solutions, extending to higher dimensions the solution \eqref{solution}-\eqref{sol4} with $b=0$. The corresponding coupling function $\epsilon(\Phi)$ will have two arbitrary couplings related to $C_1$ and $C_2$.  To be complete, note that in four dimensions, the hypergeometric function becomes constant and hence the solution is $h(x)=C_1+C_2x^{\frac{(1-4\xi)}{4\xi}}$, yielding again the neutral expression (\ref{sol4}) once expressed with the variable $r$.

To conclude, we mention that for the particular value $\xi=\frac{1}{4}$, the potential becomes
\eq
U(\Phi) =\frac{1}{2}\, \left(  \ln  \left( {\frac {\Phi}{
        b}} \right)  ^{2}+\ln  \left( {\frac {\Phi}{b}} \right)
        \left( D-1 \right) +\frac{1}{4}\, \left( D-1 \right)  \left( D-2 \right)
        \right) {\Phi}^{2},
\eeq
where $b$ is the only remaining coupling constant, and the function $\epsilon(\Phi)$ is given by
\eq
\epsilon(\Phi) =\frac{4\left(D-2\right)^{2}\left(\Phi^{2}-4\right)^{\frac{D}{D-2}}}{G(\Phi)},
\eeq
with
    \begin{eqnarray*}
        G(\Phi)=&&4\,{\Phi}^{2}
        \ln  \left( {\frac {\Phi}{b}} \right) ^{2} \left[ 4\left(D-2\right) -\left(D-4\right)\Phi^2
        \right]
        \\&&-\left( {\Phi}^{2}-4 \right) \left(D-2\right)\left(D-3\right)\left[4\,\Phi^2
        \ln  \left( {\frac {\Phi}{b}} \right)
        + \left( {\Phi}^{2}-4 \right)
        \left( D-2 \right)\right].
        \nonumber\\
    \end{eqnarray*}
In this case, the metric function $F(r)$ and the scalar field $\Phi(r)$ are given by
\begin{equation}
        F(r)={r^{2}}-\frac{p^{2}}{\left(b^{2} e^{2 a r}-4\right)^{\frac{2}{D-2}}}
        ,\qquad \Phi(r)=b e^{ar},
\end{equation}
with $a$ and $p$ being two arbitrary integration constants and where the expression of the $p-$forms is the same as the one given for $\xi\not=\frac{1}{4}$, (\ref{solaxfield}). When $D=4$, this is the double scaling limit solution \eqref{dsl14} with $\si_1=1$.


\end{document}